\documentclass[prd,twocolumn,amsmath,amssymb,floatfix, superscriptaddress,nofootinbib]{revtex4}

\usepackage{bm}
\usepackage{amsmath}
\usepackage{epsf}
\usepackage{color}
\usepackage{natbib}
\usepackage{graphicx}

\makeatletter
\renewcommand\footnoterule{%
  \vspace{-5pt}
  \kern-3\p@\hrule\@width.4\columnwidth%
  \kern10\p@}
\makeatother

\def\simlt{\lesssim}
\def\simgt{\gtrsim}
\def\be{\begin{equation}}
\def\ee{\end{equation}}
\def\ba{\begin{eqnarray}}
\def\ea{\end{eqnarray}}
\def\nn{\nonumber}
\def\Var{\mbox{Var}}

\newcommand{\lsc}{\mathcal{L}}

\newcommand{\zmax}{z_{\rm max}}
\newcommand{\Pinf}{\Delta_{{\cal R}}^2}

\newcommand{\dlnl}{-2\Delta\ln\lsc}
\newcommand{\lnl}{-2\ln\lsc}
\newcommand{\dchisq}{\Delta\chi^2}

\newcommand{\clee}{C_{\ell}^{EE}}

\def\eg{{\sl e.g.}}
\def\cf{{\sl cf.}}

\def\ie{{\sl i.e.}}

\definecolor{darkgreen}{cmyk}{0.85,0.2,1.00,0.2} 
\definecolor{purple}{cmyk}{0.5,1.0,0,0}

\begin{document}
\title{CMB polarization features from inflation versus reionization}

\author{Michael J. Mortonson}\email{mjmort@uchicago.edu}
\affiliation{Department of Physics, University of Chicago, Chicago IL 60637}
\affiliation{Kavli Institute for Cosmological Physics and Enrico Fermi Institute, University of Chicago, Chicago IL 60637, U.S.A.}

\author{Cora Dvorkin}\email{cdvorkin@uchicago.edu}
\affiliation{Department of Physics, University of Chicago, Chicago IL 60637}
\affiliation{Kavli Institute for Cosmological Physics and Enrico Fermi Institute, University of Chicago, Chicago IL 60637, U.S.A.}

\author{Hiranya V. Peiris}\email{hiranya@ast.cam.ac.uk}
\affiliation{Institute of Astronomy, University of Cambridge, Cambridge CB3 0HA, U.K.}

\author{Wayne Hu}\email{whu@background.uchicago.edu}
\affiliation{Department of Astronomy \& Astrophysics, University of Chicago, Chicago IL 60637}
\affiliation{Kavli Institute for Cosmological Physics and Enrico Fermi Institute, University of Chicago, Chicago IL 60637, U.S.A.}

\date{\today}

\begin{abstract}
\baselineskip 11pt
The angular power spectrum of the cosmic microwave background temperature anisotropy
observed by WMAP has an anomalous dip at $\ell \sim 20$ and bump at $\ell \sim 40$.
One explanation for this structure is the presence of features in the primordial curvature power
spectrum, possibly caused by a step in the inflationary potential. The detection of these features
is only marginally significant from temperature data alone. However, the inflationary feature hypothesis
predicts a specific shape for the $E$-mode polarization power spectrum with a structure similar to
that observed in temperature at $\ell \sim 20-40$.  Measurement of the CMB polarization on
few-degree scales can therefore be used as a consistency check of the hypothesis. The Planck
satellite has the statistical sensitivity to confirm or rule out the model that best fits the temperature features
with $3~\sigma$ significance, assuming all other parameters are known. With a cosmic variance limited
experiment, this significance improves to $8~\sigma$. For tests of inflationary models that can explain
both the dip and bump in temperature, the primary source of uncertainty is confusion with polarization
features created by a complex reionization history, which at most reduces the significance to $2.5~\sigma$
for Planck and $5-6~\sigma$ for an ideal experiment. Smoothing of the polarization spectrum by a
large tensor component only slightly reduces the ability of polarization to test for inflationary features,
as does requiring that polarization is consistent with the observed temperature spectrum given the
expected low level of $TE$ correlation on few-degree scales. If polarized foregrounds can be
adequately subtracted, Planck will supply valuable evidence for or against features in the primordial
power spectrum.  A future high-sensitivity polarization satellite would enable a decisive test of the
feature hypothesis and provide complementary information about the shape of a possible step in the
inflationary potential.
\end{abstract}

\maketitle

\section{Introduction} \label{sec:intro}

Our best constraints on the shape of the primordial power spectrum at large scales
come from observations of the cosmic microwave background (CMB) anisotropy 
by the Wilkinson Microwave Anisotropy Probe (WMAP) \cite{Bennett:2003bz,Hinshaw:2008}. 
The latest (5-year) WMAP data \cite{Nolta:2008ih, Dunkley:2008ie} 
continue to be well described by the simplest inflationary scenario of a 
single, slowly rolling, minimally coupled scalar field with a canonical kinetic term 
\cite{Komatsu:2008hk,Peiris:2008be, Kinney:2008wy,Alabidi:2008ej,Lesgourgues:2007gp}. 
Since the 3-year release \cite{Spergel:2006hy}, the WMAP data have indicated a deviation 
from scale invariance --- a red tilt of the scalar spectral index --- the significance of which 
has been debated in the literature from a Bayesian model selection point of view 
(\eg\ \cite{Parkinson:2006ku, Gordon:2007xm}). Recent minimally-parametric reconstructions
of the primordial power spectrum incorporating some form of penalty for ``unnecessary''
complexity \cite{Verde:2008, Bridges:2008ta} show some evidence for a red tilt, but 
no evidence for scale dependence of the spectral index. These methods,
as currently implemented, are not very sensitive to sharp, localized features in the primordial power
spectrum.

However, it has been pointed out ever since the original data release \cite{Spergel:2003cb, 
Peiris:2003ff} that there are several sharp glitches in the WMAP temperature ($TT$) power spectrum. 
In particular, several model-independent reconstruction techniques that are sensitive to 
features localized in a narrow wavenumber range 
have consistently picked out a feature at 
$\ell \sim 20-40$ that leads to an improvement of $\dchisq \sim {\cal O}(10)$ over a
smooth power-law spectrum \cite{Hannestad:2003zs, Shafieloo:2003gf, 
Mukherjee:2003ag, Shafieloo:2006hs, Nicholson:2009pi}. 

Power spectrum features could arise, in principle, in more general classes of inflationary models 
where slow roll is momentarily violated. Such an effect can be phenomenologically modeled
as a discontinuity or singularity in the inflaton potential \cite{Starobinsky:1992ts, Adams:2001vc, Joy:2007na}. 
A ``step-like'' feature \cite{Adams:2001vc}, in particular, would be a good effective field theory
description of a symmetry breaking phase transition in a field coupled to the inflaton in
multi-field models \cite{Silk:1986vc, Holman:1991ht, Polarski:1992dq, 
Adams:1997de, Hunt:2004vt}, which can arise in supergravity \cite{Lesgourgues:1999uc} or 
M-theory-inspired \cite{Burgess:2005sb,Ashoorioon:2006} contexts. Several analyses have confronted such 
phenomenological descriptions of features in the inflationary potential with current data 
\cite{Peiris:2003ff, Covi:2006ci, Hamann:2007pa, Hunt:2007dn, Joy:2008qd, Martin:2003sg, Kawasaki:2004pi, Jain:2008dw}.

It is debatable whether the large scale feature seen in the WMAP $TT$ spectrum is
a signal of exotic primordial physics or merely a statistical anomaly. Currently,
our information about the smoothness of the primordial power spectrum is dominated by 
the temperature data. However, future high fidelity CMB polarization measurements at 
large scales have the potential to shed light on this question. The importance of 
polarization data for constraining oscillatory features has been previously discussed in 
the literature (\eg\ \cite{Hu:2003vp, Kogo:2004vt, Pahud:2008, Nagata:2008zj, Nicholson:2009pi}) 
and exploited in particular as a cross-check of the observed low CMB temperature quadrupole 
\cite{Dore:2003wp,Skordis:2004xr,GorHu04,Fanetal08}. 

In this work, we propose to use the large-scale polarization of the CMB to test 
the hypothesis that the $\ell \sim 20-40$ glitch is due to a step 
in the inflaton potential. We exploit the fact that, in the relevant multipole range, 
the sharpness of the polarization transfer function and 
lack of contamination by secondary effects (assuming 
instantaneous reionization) makes polarization a cleaner probe of 
such features than temperature \cite{Hu:2003vp}.
We also investigate how our conclusions
are affected by relaxing the assumption of instantaneous reionization
\cite{Hu:2003gh,Mortonson:2007hq,Mortonson:2007tb}, changing the parameters of the
 feature, and 
including large-amplitude tensor fluctuations. This analysis
is particularly timely given the imminent launch of the Planck satellite \cite{Planck:2006uk},
which promises to greatly increase our knowledge of the large-scale polarization signal. It is 
also relevant for future dedicated CMB polarization missions \cite{Baumann:2008aq}.
As in a related previous paper on polarization consistency tests of large-angle CMB
temperature anomalies \cite{Dvorkin:2007}, it is our objective to make a prediction
for the polarization statistics that will be observed by future CMB experiments, given current
temperature data, in a ``last stand" before Planck.

We present the inflationary model and the numerical procedure used to compute the primordial 
curvature power spectrum in \S~\ref{sec:model}. The polarization consistency tests of the features, both for
instantaneous and general reionization histories, are presented in \S~\ref{sec:pol_tests} and \S~\ref{sec:ionization}, and we 
conclude in \S~\ref{sec:discuss}.
We discuss in Appendix~\ref{sec:appendix} the relation of our work
to previous analyses of features in the WMAP temperature data.

\section{Inflationary Features} \label{sec:model}

We review the inflationary generation of features in the curvature power spectrum
from step-like features in the inflaton potential in \S \ref{sec:inflation} and their transfer to the
CMB temperature and power spectra in \S \ref{sec:cmb}.

\subsection{Inflationary Model}
\label{sec:inflation}

To model a feature in the primordial power spectrum that matches the 
glitches in the WMAP temperature data at $\ell \sim 20-40$, 
we adopt a phenomenological
inflationary potential of the form $V(\phi)=m_{\rm eff}^2(\phi)\phi^2/2$ 
where the effective mass of the inflaton $\phi$ has a step at 
$\phi = b$ corresponding to the sudden change 
in mass during a phase transition \citep{Adams:2001vc}:
\begin{equation}
m_{\rm eff}^2(\phi) = m^2 \left[1+c\tanh\left(\frac{\phi-b}{d}\right)\right],
\label{eq:meff}
\end{equation}
with the amplitude and width of the step determined by $c$ and $d$ respectively, assuming that both are positive numbers. 
We express the potential parameters $m$, $b$, and $d$ in units of the 
reduced Planck mass, $M_{\rm Pl} = (8\pi G)^{-1/2} = 2.435\times 10^{18}$~GeV; 
the step amplitude $c$ is dimensionless.

In physically realistic models with a sufficiently small step in the potential, 
the interruption of slow roll as the field encounters 
the step does not end inflation but affects density perturbations
through the generation of scale-dependent oscillations that eventually die away. 
The phenomenology of these oscillations is described in Ref. \cite{Adams:2001vc}:
the sharper the step, the larger the amplitude and width of the ``ringing''
superimposed upon the underlying smooth power spectrum. 
Hence we shall see in \S~\ref{sec:inflationarydegradation} that lowering $d$ increases the
width of the feature in $\ell$ in the CMB power spectra.

Standard slow-roll based approaches are
insufficient for computing the power spectrum for this potential, 
and instead the equation of motion must be integrated numerically
mode-by-mode \cite{Leach:2000yw}. It is convenient to use the gauge invariant
Mukhanov potential \cite{mukhanov:1988, sasaki:1986} for the mode amplitude
since it is simply related to the curvature 
perturbation ${\cal R}$: 
\begin{equation}
u = -z {\cal R}\,, 
\end{equation}
where $z\equiv \dot{\phi}/H$, $H$ is the Hubble
parameter, and the dot denotes a derivative with respect to conformal time. The Fourier components $u_k$
obey the equation of motion \cite{stewart:1993, mukhanov:1985,mukhanov:1992}
\begin{equation}
\ddot u_k + \left( k^2 - \frac{\ddot z}{z}\right) u_k = 0\,, 
\label{eq:mode}
\end{equation}
where $k$ is the
modulus of the wavevector ${\bf k}$.  The power spectrum is 
defined via the two point correlation function
\begin{equation}
\langle {\cal R}_{{\bf k}\vphantom{'}} {\cal R}^*_{{\bf k}'} \rangle = \frac{2\pi^2}{k^3} \Delta_{\cal R}^2(k)\
(2\pi)^3 \delta^{(3)}({\bf k} -{\bf k}'),
\end{equation}
which is related to $u_k$ and $z$ via 
\begin{equation}
\Delta^2_{\cal R}(k) = \frac{k^3}{2\pi^2}\left| \frac{u_k}{z}\right|^2.
\end{equation}

The dynamics of the Hubble parameter, described by the Friedmann equation, and the background 
dynamics of the unperturbed inflaton field, described by the Klein-Gordon equation, can be written 
respectively as
\begin{flalign}
& H' = -\frac{1}{2} H (\phi')^2\ ,\label{eq:evol1}\\
& \phi'' + \left( \frac{H'}{H} + 3 \right)\phi' + \frac{1}{H^2} \frac{dV}{d\phi} = 0\,,\label{eq:evol2}
\end{flalign}
where $'=d/d\ln a$.
The solution of the mode equation depends on the background  dynamics.
With the help of these background equations, the mode equation~(\ref{eq:mode}) can be written as
\begin{flalign}
& u_{k}'' + \left(\frac{H'}{H} + 1 \right)u_{k}' + \left\{\frac{k^2}{a^2 H^2} -  \Bigg[ 2  - 4\,\frac{H'}{H} \frac{\phi''}{\phi'}  \right.   \nonumber \\
& \quad \left.\left.  - 2 \left( \frac{H'}{H}\right)^2  - 5 \frac{H'}{H}  - \frac{1}{H^2} \frac{d^2 V}{d\phi^2} \right] \right\}u_k = 0 \,,
\label{eq:evol3}
\end{flalign}
where the term in square brackets is $\ddot z/(za^2H^2)$.
We stop integrating the background equations after any transient solution has died away, 
but while the mode is still well within the horizon. This allows us to obtain initial conditions for the
two orthogonal solutions that contribute to $u_k$ that are free of contamination due
to any transient contribution to the background dynamics. The power spectrum is
then obtained by continuing the integration until the mode freezes out far outside the
horizon, yielding the asymptotic value of $|u_k/z|$.  
Further details regarding 
the numerical solution of the coupled system of differential equations can be found 
in Ref. \cite{Adams:2001vc}.

\begin{figure}[tb]
  \resizebox{3.4in}{!}{\includegraphics[angle=0]{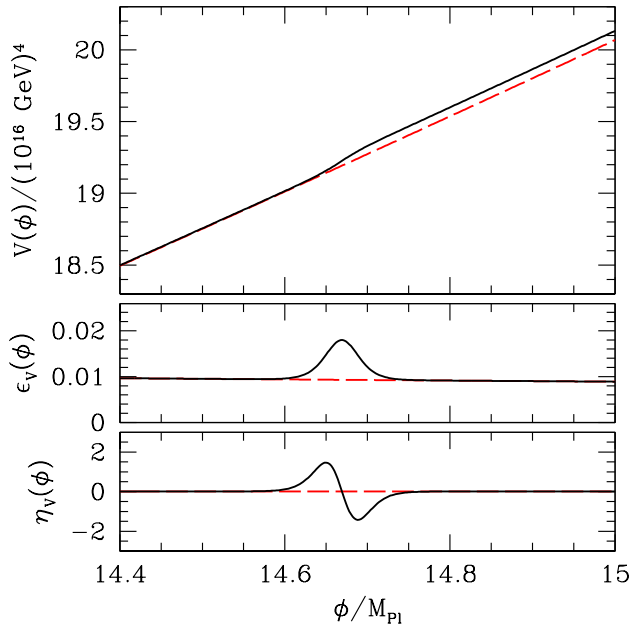}}
  \caption{\footnotesize
\emph{Upper panel, solid black}: 
Inflationary potential with a step [Eq.~\ref{eq:meff}].
The parameters for the potential are chosen to maximize the 
WMAP5 likelihood and are listed in Table~\ref{tab:modelparameters}. 
The dashed red line shows a smooth $m^2\phi^2$ potential ($c=0$) with 
$m=7.120\times 10^{-6}$ so that the two models have equal power 
on small scales ($\phi\ll b$).
\emph{Middle and lower panels}: slow-roll parameters $\epsilon_V$ and $\eta_V$
for the two inflationary potentials.
  }
  \label{plot:potential}
\end{figure}

To match a given mode to a physical wavenumber $k$, one must make an assumption 
about the reheating temperature, but this choice is degenerate with $b$, corresponding to 
a translation of the step in $\phi$. To compare our results with
those of Refs.~\cite{Covi:2006ci, Hamann:2007pa}, we adopt the following prescription
for the matching:
\begin{equation}
k_\star \equiv a_\star H_\star = a_{\rm end} e^{-N_\star} H_\star,
\label{eq:kmatch}
\end{equation}
where $H_\star$ is the Hubble scale corresponding to the physical wavenumber $k_\star$,
which left the horizon $N_\star$ $e$-folds before the end of inflation, defined by $d^2 a/dt^2(a_{\rm end})=0$.
Following 
the above authors, we set the pivot scale $k_\star=0.05$ Mpc$^{-1}$ to correspond to $N_\star=50$ (although there are differences in the implementation of 
the $k$-mode matching that we discuss in Appendix~\ref{sec:appendix}).

\begin{table}
\caption{Fiducial feature model parameters chosen to best fit WMAP5 under a 
flat $\Lambda$CDM cosmology, compared in the text with a smooth model with
$c=0$, $m=7.120 \times 10^{-6}$, and the same cosmological parameters, which
matches the small scale normalization $A_s(k_\star)=2.137\times 10^{-9}$ and 
tilt $n_s \approx 0.96$ at the pivot $k_\star = 0.05$ Mpc$^{-1}$.}
\begin{center}
\begin{tabular}{cc}
\hline
Parameter & Value\\
\hline
\hline
$m$ & $7.126 \times 10^{-6}$ \\
$b$  & $14.668$\\
$c$  & $1.505 \times 10^{-3}$ \\
$d$ & 0.02705  \\
$N_\star$ & 50 \\
\hline
$\Omega_b h^2$ & 0.02238 \\
$\Omega_c h^2$ & 0.1081\\
$h$ & 0.724\\
$\tau$ & 0.089\\
\hline\\
\end{tabular}
\end{center}
\label{tab:modelparameters}
\end{table}

\begin{figure}[tb]
  \resizebox{3.4in}{!}{\includegraphics[angle=0]{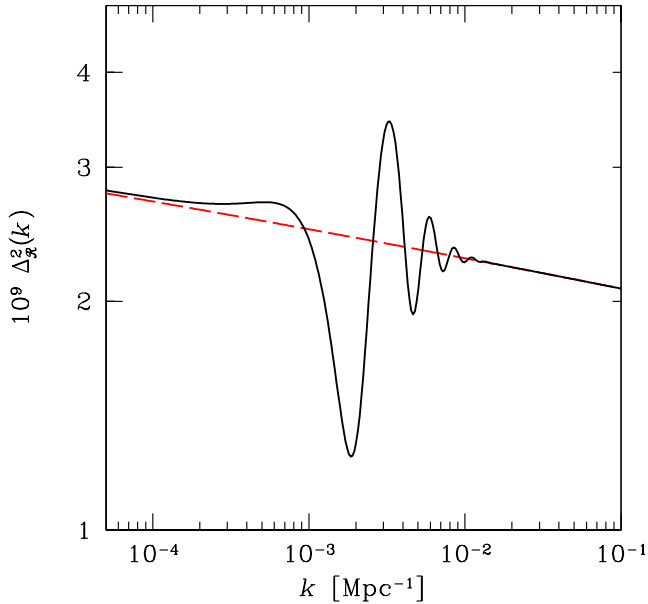}}
  \caption{\footnotesize
Primordial curvature power spectra for the potentials in Fig.~\ref{plot:potential}. 
  } 
  \label{plot:powspec}
\end{figure}

Figure~\ref{plot:potential} shows our fiducial inflationary potential, 
with parameters given in Table~\ref{tab:modelparameters} that are chosen to 
fit the WMAP5 temperature glitches at $\ell\sim 20-40$ as we will 
show in the next section. 
The number of $e$-folds of 
inflation after the step in this potential is $N_{\rm step}\approx 54$.
The slow-roll parameters
\begin{equation}
\epsilon_V=\frac{M_{\rm Pl}^2}{2}\left(\frac{dV/d\phi}{V}\right)^2,\quad
\eta_V=M_{\rm Pl}^2 \frac{d^2V/d\phi^2}{V}
\end{equation}
are plotted in the lower panels of Fig.~\ref{plot:potential}. Note that 
near the step at $\phi=b$, $|\eta_V| \gtrsim 1$ confirming that the
slow-roll approximation is not valid.
Figure~\ref{plot:powspec} shows the inflationary curvature power spectrum 
$\Pinf(k)$ for this potential, computed 
by integrating Eqs.~(\ref{eq:evol1})$-$(\ref{eq:evol3}).

For comparison, in Figs.~\ref{plot:potential} and~\ref{plot:powspec} we also 
show a smooth, $c=0$ potential with the same small-scale amplitude and tilt 
as the fiducial potential, and its slow-roll parameters and inflationary 
power spectrum.
The smooth spectrum is nearly indistinguishable
from a pure power law of $n_s \sim 0.96$ with amplitude
$A_s(k_\star)=2.137\times 10^{-9}$.
Note that the spectral index is determined by the choices of 
$N_{\star}$ and $k_{\star}$ in the matching condition 
of Eq.~(\ref{eq:kmatch}), while the amplitude comes from 
the inflaton mass $m$.

\subsection{CMB Power Spectra}
\label{sec:cmb}

The mapping between the inflationary curvature power spectrum and the
observable CMB angular power spectra
\begin{equation}
\langle X_{\ell m}^* X'_{\ell' m'} \rangle = \delta_{\ell\ell'}\delta_{mm'} 
	C_\ell^{XX'}\,,
\label{eqn:obspowerdef}
\end{equation}
where $X,X' \in T,E$, is given
by the scalar radiation transfer functions
\begin{equation}
{\ell (\ell+1) C_\ell^{XX'} \over 2\pi} = \int {d \ln k} \,
T^{X}_\ell(k) T^{X'}_\ell(k)\,
\Pinf(k)\,.
\label{eqn:transdef}
\end{equation}

\begin{figure}[tb]
  \resizebox{3.2in}{!}{\includegraphics[angle=0]{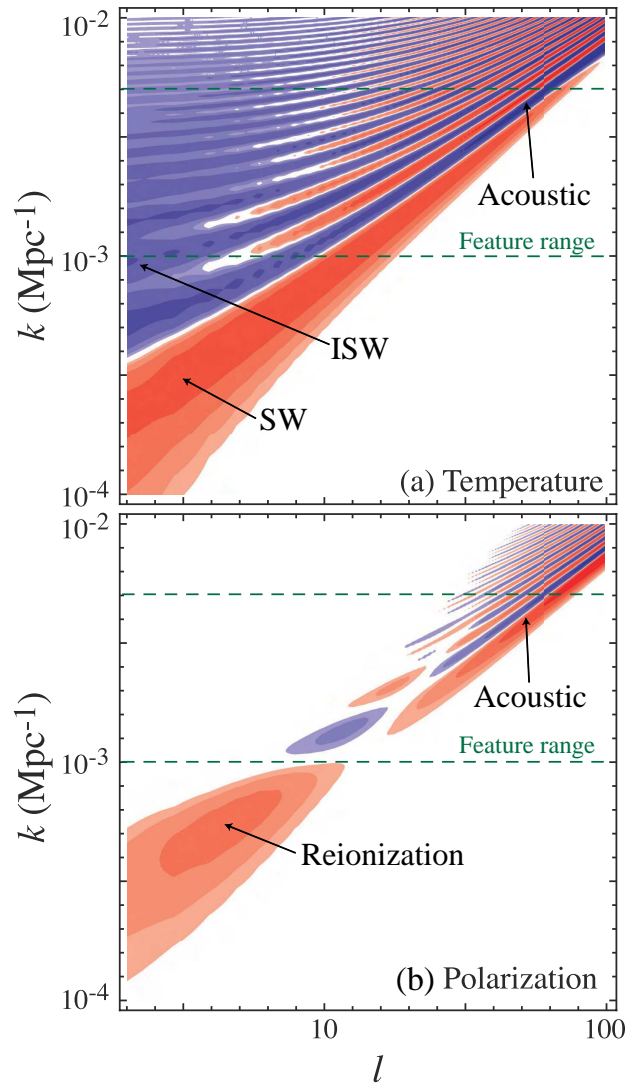}}
  \caption{\footnotesize
  Transfer function $T_\ell^X(k)$ for the fiducial model with instantaneous reionization.
  \emph{Upper panel:} temperature $X=T$; \emph{lower panel:} polarization $X=E$.  Contours are spaced by 
  factors of $2$.  Dashed lines
  represent the range of $k$-modes where features appear in Fig.~\ref{plot:powspec}. Polarization is
  a cleaner probe of features in this range and, for instantaneous reionization, is nearly  uncontaminated
  by secondary effects.   The temperature and polarization are also only weakly correlated
  here due to the transition between the Sachs-Wolfe (SW) and acoustic regimes in temperature.
}
  \label{plot:transfer}
\end{figure}

\begin{figure}[tb]
  \resizebox{3.2in}{!}{\includegraphics[angle=0]{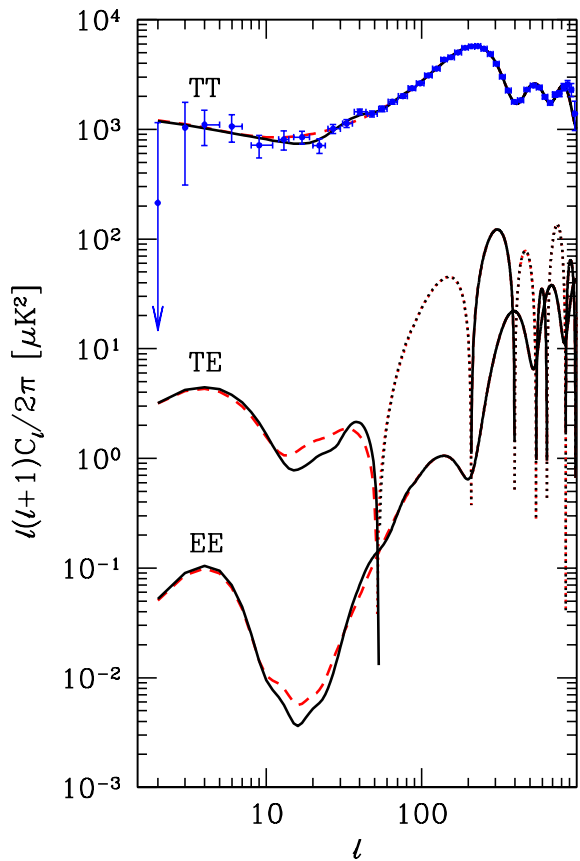}}
  \caption{\footnotesize
Temperature and polarization power spectra for the inflationary 
power spectra in Fig.~\ref{plot:powspec}, with solid black curves 
for the model with a feature and dashed red curves for 
smooth $\Pinf(k)$. Dotted curves indicate where $C_{\ell}^{TE}$ is negative.
Blue points with error bars show the 5-year WMAP measurements 
of $C_{\ell}^{TT}$ including sample variance.
For both models, 
the reionization history is assumed to be instantaneous and 
the cosmological parameters not determined by the inflationary 
potential are given in Table~\ref{tab:modelparameters}.  
  } 
  \label{plot:cltt}
\end{figure}

In Fig.~\ref{plot:transfer}, we show the $T$ and $E$ transfer functions for the
fiducial cosmological parameters of
Table~\ref{tab:modelparameters}.  For a more extended discussion of the 
transfer functions and their relationship to features in the inflationary power spectrum,
see \cite{Hu:2003vp}. 
The resultant temperature and polarization 
angular power spectra from the inflationary power spectra of Fig.~\ref{plot:powspec}
are 
plotted in Fig.~\ref{plot:cltt}.

\begin{figure}[tb]
  \resizebox{3.4in}{!}{\includegraphics[angle=0]{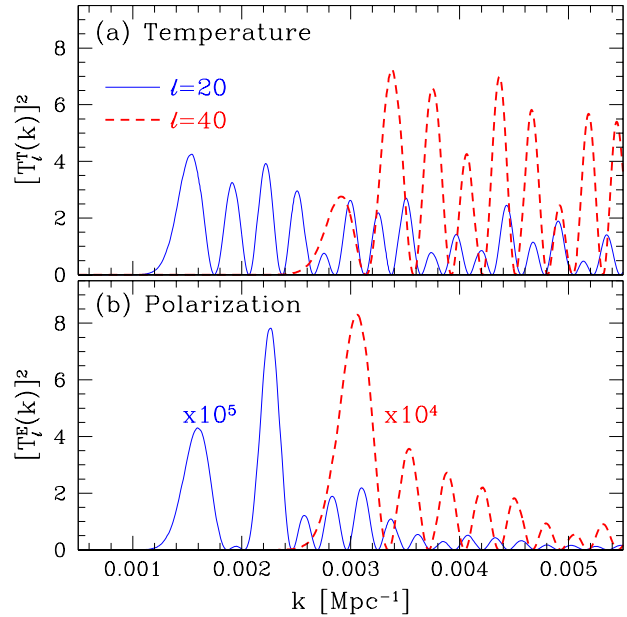}}
  \caption{\footnotesize
  Transfer function $T_\ell^X(k)$ for the fiducial model with instantaneous reionization for
  multipoles near the temperature dip ($\ell = 20$) and bump ($\ell = 40$) for
  temperature and polarization.   For temperature, the dip multipoles receive a broad
  range of contributions from $k \simgt 10^{-3}$ and the bump multipoles from $k \simgt 3
  \times 10^{-3}$.  The localization of the transfer function is sharper 
for polarization, especially
  for $\ell = 40$ which is immune to reionization effects.
   The polarization transfer functions have been scaled by $10^4$ and $10^5$ for
   convenience.
}
  \label{plot:tkl}
\end{figure}

  For the wavenumbers of interest,
$1 \simlt k/10^{-3} {\rm Mpc}^{-1} \simlt 5$, the transfer of power to temperature fluctuations 
transitions between the Sachs-Wolfe and acoustic regimes at high $\ell$ and 
carries substantial contributions from the integrated Sachs-Wolfe (ISW) 
effect at low $\ell$.   These effects and geometric
projection lead to
a very broad mapping of power in $k$ to power in $\ell$.    In particular, the oscillations at
the upper range in $k$ are largely washed out, leaving only a single
broad dip at $\ell\sim 20$ and bump at $\ell\sim 40$ in the 
temperature spectrum.  Likewise,
the power at these multipoles correspond to a wide range in $k$ as shown in Fig.~\ref{plot:tkl}.

Polarization spectra differ notably from the temperature spectra due to 
the differences in the transfer function shown in Fig.~\ref{plot:transfer}.
  For the standard instantaneous reionization
history and the upper portion of the range of $k$ affected by the feature, the polarization is dominated by the onset of acoustic effects only.  We shall see that this makes the
bump in $\ell\sim 40$ a particularly clean test of inflationary features (see Fig.~\ref{plot:tkl}).  
Furthermore oscillations from high $k$ at higher $\ell$ are retained at a
significant level in the polarization.  

On the other hand at $k \sim 10^{-3}$ Mpc$^{-1}$, the polarization transfer from 
recombination becomes very inefficient and
reionization effects come into play.   This leads to a very low level of
polarization around $\ell \sim 20$ with features even for a smooth inflationary 
power spectrum.   These properties leave the $\ell \sim 20$ dip vulnerable to 
external contamination such as tensor contributions (see \S~\ref{sec:tensors}) or foregrounds
as well as uncertainties in the ionization history (see \S~\ref{sec:ionization}).

Finally,  the cross correlation between the
temperature and polarization fields for the entire range of $20 \le \ell \le 40$ is very low due to the transition
between the Sachs-Wolfe and acoustic-dominated regimes in the temperature field.
We shall see in \S~\ref{sec:constrained} that this prevents statistical fluctuations in the
observed temperature power spectrum from being repeated in the polarization.

These differences in the transfer functions also play a role in defining the
region in the potential parameter space that best fits the WMAP $TT$ data versus 
the region that is best
tested by polarization.   For the former, we conduct a grid based search over
the potential parameters.
The mass parameter $m$ determines the amplitude of the spectrum away from the feature 
and so is mainly fixed by the acoustic peaks at high $\ell$.  
The location of the feature $b$ is also well determined independently of the other parameters~\cite{Covi:2006ci, Hamann:2007pa}.
We therefore fix $m$ and $b$ at their best-fit values and search for the 
best fit in the step amplitude and width parameters $c$ and $d$.

The values of $m$, $b$, $c$, and $d$
given in Table~\ref{tab:modelparameters} specify the maximum likelihood
model.  This model improves the fit to the 5-year WMAP data by 
$\dlnl_{TT}\approx -8$.  
  We will explore variations in the parameters about the maximum and their
relationship to the temperature and polarization power spectra through the transfer
functions in \S~\ref{sec:inflationarydegradation}.  

 The improvement is only 
marginally significant given the 3 extra parameters of the step and the choice of one out of many possible forms. 
 Matching polarization features can therefore provide a critical confirmation or
refutation 
of the inflationary nature of the temperature features.

\section{Confirming Features with Polarization} \label{sec:pol_tests}

In this section, we discuss the significance with which
polarization measurements can confirm or rule out the inflationary features 
discussed in the previous section under the instantaneous reionization model.
We begin in \S~\ref{sec:basesignificance} with the significance of the 
best-fit feature model under the simplest set of assumptions.
We assess changes in the significance 
due to variation in the potential parameters in \S~\ref{sec:inflationarydegradation}, and due to the inclusion of tensor $E$-modes in \S~\ref{sec:tensors}.
In \S~\ref{sec:constrained}, we describe the impact of conditioning
polarization predictions on the already-measured temperature spectrum.

\subsection{Fiducial Polarization Significance}
\label{sec:basesignificance}

To evaluate the significance of discriminating between models, we 
assume a Gaussian likelihood for the polarization angular power spectrum.
In the absence of detector noise,
the likelihood $\lsc_{EE}$ of data $\hat{C}_{\ell}^{EE}$ given a model 
power spectrum $C_{\ell}^{EE}$ is 
\begin{equation}
\lnl_{EE} \approx f_{\rm sky} \sum_\ell (2\ell +1) \left(
{\hat C_\ell^{EE} \over C_\ell^{EE} }+ \ln {C_\ell^{EE} \over \hat C_\ell^{EE}} -1 \right) \,,
\label{eq:likelihood}
\end{equation}
where $f_{\rm sky}$ is the fraction of sky with usable $E$ measurements.
If the data $\hat{C}_{\ell}^{EE}$ have no inflationary feature and the 
model $\clee$ has the inflationary feature, we call
this the significance at which {\it false positives} can be rejected.   
Conversely, if the data
have an inflationary feature and the model spectrum has no feature, 
we call this the significance
at which {\it false negatives} can be rejected.

\begin{table}
\caption{Parameters used when making forecasts for idealized 
and Planck-like experiments. Here $\Delta_P^{(\nu)}$ is in units of $\mu$K-arcmin.}
\begin{center}
\begin{tabular}{ccccc}
\hline
Experiment &  $\nu$  & $\theta_{\rm FWHM}^{(\nu)}$ & $\Delta_P^{(\nu)}$  & $f_{\rm sky}$   \\
\hline
\hline
 Ideal & --- & 0 & 0 & 0.8 \\
\hline
 Planck  &    $70$ GHz    &  $14.0'$ & $255.6$& $0.8$  \\
&$100$ GHz & $10.0'$ & $109.0$ & $0.8$  \\
&$143$ GHz & $7.1'$ & $81.3$  & $0.8$  \\
\hline
\end{tabular}
\end{center}
\label{tab:Planck-specs}
\end{table}

For forecasts throughout this paper, we assume that 
the data are equal to the ensemble average of realizations for a particular 
model. Therefore, the minimum $\lnl$ is zero and 
$\dlnl = \lnl$. 
The exception to this is that when we discuss the 
WMAP $TT$ likelihood, the relevant quantity is $\dlnl = -2\ln(\lsc/\lsc_{\rm ML})$ 
where the likelihood of the best fit model is $\lnl_{\rm ML}\ne 0$.
We make forecasts for an ideal, sample variance limited experiment and for Planck using the experimental specifications in Table~\ref{tab:Planck-specs}.
For the Planck case with a finite noise power
$N_\ell$, $C_\ell \rightarrow C_\ell + N_\ell$ in Eq.~({\ref{eq:likelihood}),
where $N_\ell$ is the minimum variance combination of the noise powers
of the individual frequency channels
\begin{equation}
N_{\ell}^{(\nu)}=\left(\frac{\Delta_P^{(\nu)}}{\mu\text{K-rad}}\right)^2
\exp\left[\frac{\ell (\ell+1)(\theta_{\rm FWHM}^{(\nu)}/{\rm rad})^2}{8\ln 2}\right]
\end{equation}
using $\Delta_P^{(\nu)}$ and $\theta_{\rm FWHM}^{(\nu)}$ from Table~\ref{tab:Planck-specs}
converted to the appropriate units.

\begin{table}
\caption{$\dlnl_{EE}$ for false positive and false negative tests comparing models with smooth $\Pinf(k)$ and a feature in $\Pinf(k)$.}
\begin{center}
\begin{tabular}{llr}
\hline
Experiment & Test & $\dlnl_{EE}$ \\
\hline
\hline
Ideal & False positive &  $64$ \\
Ideal & False negative &  $60$ \\
\hline 
Planck & False positive &  $8$ \\
Planck & False negative &  $9$ \\
\hline
\end{tabular}
\end{center}
\label{table:instrei1}
\end{table}

Table~\ref{table:instrei1} lists $\dlnl_{EE}$ for rejecting false positives 
and false negatives. 
The significance of false positive or negative rejection in this 
most optimistic case is $\sqrt{\dlnl_{EE}}\sim 8$ for the ideal 
experiment and $\sim 3$ for Planck.
In the following sections, we will discuss various effects that 
can degrade this significance.

\subsection{Potential Parameters}
\label{sec:inflationarydegradation}

Variation in the parameters of the inflationary potential from 
the best fit model can affect the significance of polarization 
tests of features. As noted in \S~\ref{sec:cmb}, $m$ and $b$ are 
strongly constrained by the observed CMB temperature spectrum, 
but the parameters $c$ and $d$ that control the amplitude and 
width of an inflationary step are less well determined by 
temperature alone.

In terms of the curvature power spectrum, increasing $c$ increases the amplitude
of the features. 
However, decreasing the width of the 
potential step by lowering $d$ enhances the deviations from slow roll,
thereby also amplifying the feature in the power spectrum.

\begin{figure}[b]
  \resizebox{3.4in}{!}{\includegraphics[angle=0]{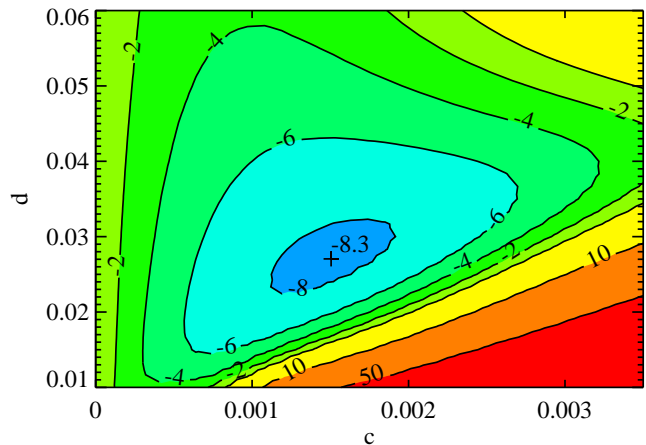}}
\caption{\footnotesize
Contour plot of $\dlnl_{TT}$ for parameters $c$ and $d$ using 
$5$-year WMAP data. Other potential parameters are fixed at their fiducial values.
The minimum,
with $\dlnl_{TT}=-8.3$ relative to the smooth $c=0$ model, 
is shown with a cross.
}
\label{plot:contour_plot_TT_WMAP5}
\end{figure}

\begin{figure}[tb]
  \resizebox{3.4in}{!}{\includegraphics[angle=0]{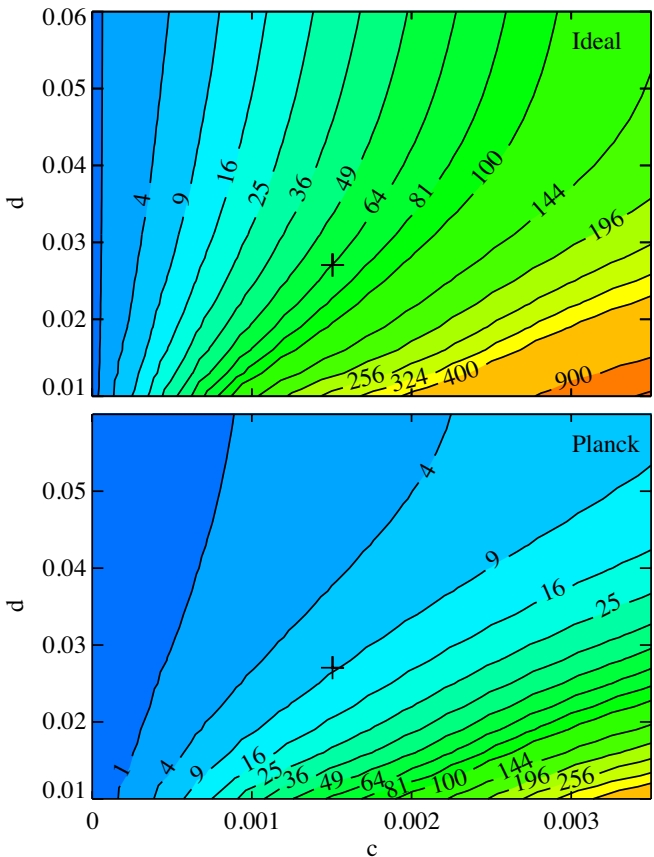}}
\caption{\footnotesize
Contour plot of $\dlnl_{EE}$ for the parameters $c$ and $d$, 
for tests of false positives with 
a cosmic variance limited experiment (\emph{upper panel}) and Planck (\emph{lower panel}). 
The best fit model to WMAP $TT$ is shown with a cross.}
\label{plot:contour_plot_EE_CV}
\end{figure}

Figure~\ref{plot:contour_plot_TT_WMAP5} shows a contour plot of 
the WMAP temperature likelihood $\dlnl_{TT}$ for the 
parameters $c$ and $d$ (relative to $c=0$) and 
Fig.~\ref{plot:contour_plot_EE_CV} shows 
$\dlnl_{EE}$ for false positives using simulated polarization data.  
The similarities and differences
between these two plots reflect properties of the temperature and polarization
 transfer functions.

For the temperature case
near the minimum, the
degeneracy between the two parameters is approximately $c\propto d^2$.  This line roughly corresponds to  
keeping the amplitude of the enhanced power in $\Pinf(k)$ at $k \sim 3 \times 10^{-3}$~Mpc$^{-1}$ fixed.
The preferred value of  $c$ corresponds to the best amplitude of the negative
dip at $k \sim 2 \times 10^{-3}$~Mpc$^{-1}$.
For the best fit parameters, including the feature in 
$\Pinf(k)$ improves the fit to 5-year WMAP data by $\dlnl_{TT} \approx -8$.

Due to the weak significance of the feature detection, the contours 
become substantially distorted away
from the maximum likelihood.   In particular, the contours in Fig.~\ref{plot:contour_plot_TT_WMAP5}
show a triangular region extending to high $d\sim 0.04$.
 This region corresponds to a lower amplitude in {\it both} the first dip and bump in $k$
 as shown in Fig.~\ref{plot:models_Pofk}. 
Due to projection effects in temperature, the $\ell = 20$ dip gets contributions from both
the dip and the bump in $k$ (see Fig.~\ref{plot:tkl}).   Consequently, a model with smaller features in $k$
in both the dip and bump can lead to the {\it same} amplitude of the dip at $\ell = 20$ if the
amplitude of the bump is reduced more.

\begin{figure}[tb]
  \resizebox{3.4in}{!}{\includegraphics[angle=0]{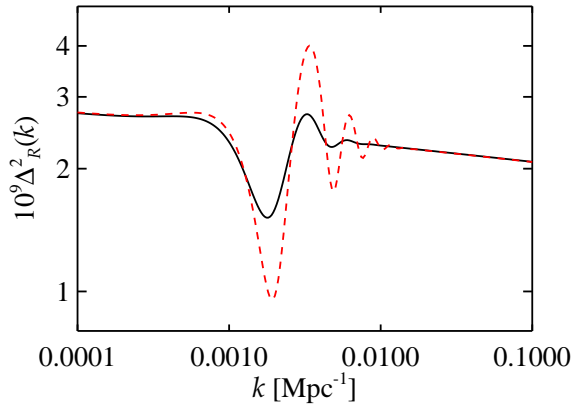}}
\caption{\footnotesize Primordial curvature power spectra for models illustrating 
projection degeneracies in the temperature.  The parameters of the two models are chosen to have
similar temperature dips at $\ell \sim 20$ and equal WMAP likelihoods:
$(c,d)=(0.00128,0.043)$ (\emph{solid black}) and $(c,d)=(0.0023,0.028)$ (\emph{dashed red}); other parameters are fixed to the values in Table~\ref{tab:modelparameters}
for both models.}
\label{plot:models_Pofk}
\end{figure}

\begin{figure}[tb]
  \resizebox{3.4in}{!}{\includegraphics[angle=0]{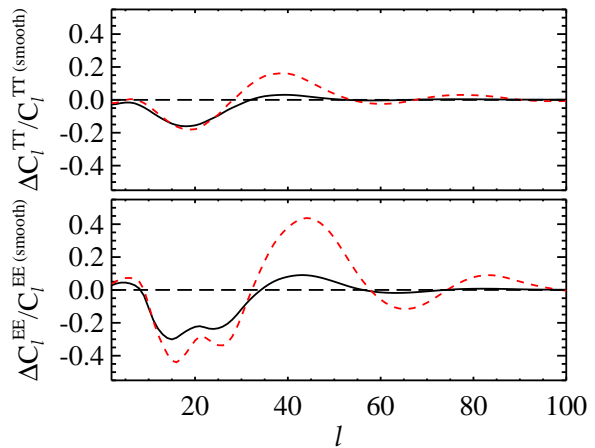}}
\caption{\footnotesize \emph{Upper panel:} Relative difference in
$C_{\ell}^{TT}$,  with respect to a smooth power spectrum, of
 two models with equal $TT$ amplitude at $\ell \sim 20$ and curvature power spectra
 shown in Fig.~\ref{plot:models_Pofk}. 
\emph{Lower panel:} Due to differences in the polarization 
transfer function, the models do not have degenerate $\ell \sim 20$ dips and show
significant differences at $\ell \simgt 60$ as well, leading to a much higher 
polarization significance
for the model plotted with dashed lines.}
\label{plot:models_powersp}
\end{figure}

We illustrate these projection effects 
in Figs.~\ref{plot:models_Pofk} and~\ref{plot:models_powersp} with two models
chosen to have the same likelihood improvement of $\dlnl_{TT} \approx -6$.
For the model with smaller features in $k$, the temperature 
enhancement at $\ell \sim 40$ is substantially reduced compared with 
the best-fit model, while the model with larger features in $k$ overshoots the 
bump at $\ell=40$ in temperature. Despite these differences,
both models have about the same $TT$ amplitude in the $\ell=20$ dip
as the best-fit model but a slightly worse overall fit.
In particular, for the model with smaller features in $k$, inflationary features
can
only explain the observed $\ell=20$ dip in temperature and not the
$\ell=40$ bump.

The degeneracies in $c$ and $d$ for polarization significance share similarities with, yet have
important differences from, those for temperature.
The polarization significance remains largely unchanged for
small variations in $c$ and $d$  along the constant $c/d^2$ line favored by the temperature spectra.
Near the maximum, variations along this direction
preserve the amplitude of intrinsic features in $k$
(see Fig.~\ref{plot:contour_plot_TT_WMAP5}).  However within the $\dlnl_{TT} =-4$ region the
significance for a ideal experiment can either drop or rise significantly.  
The reason is that due to projection effects in temperature, the polarization better separates
changes in the overall and relative amplitude of the features in $k$. 
  In the triangular high $d$ region,
where the amplitude of the $TT$ dip remains unchanged but the intrinsic features in
$k$ are all reduced, the significance of the polarization difference decreases markedly
(see Fig.~\ref{plot:models_powersp}).
Because of the sharper projection, even the $\ell \sim 20$ dip in polarization is reduced. 
The net result is that the polarization significance is a stronger function of $c$, which 
controls 
the overall amplitude, than the temperature significance.

Note that while the significance can be substantially degraded from our best fit assumptions,
this is mainly because
of the weak detection of a feature in the temperature spectrum itself.  In cases where
the polarization significance is greatly reduced, the temperature bump at $\ell \sim 40$
cannot be explained by the inflationary features.
In other words,
polarization remains a robust probe of the inflationary nature of the $\ell = 40$ bump
across variations of the potential parameters.

\subsection{Tensors}
\label{sec:tensors}

The $m^2 \phi^2$ potential with the parameters in Table~\ref{tab:modelparameters}
predicts substantial gravitational wave contributions
with tensor-to-scalar ratio $r\approx 0.16$. 
Relative to a smooth-$\Pinf(k)$ model
without tensors, the $m^2 \phi^2$ model with a feature has extra distinguishing power
due to the presence of $B$-mode polarization. 
 Because other forms for the  potential can also be used as the
smooth base on which to place the feature
\cite{Hamann:2007pa}
we choose not to include tensors for most of our calculations.   Moreover, a $B$-mode detection
would not be useful for discriminating features. 
The $B$-mode amplitude is insensitive to features 
since the potential amplitude is left nearly unchanged by the step.
Additionally, a small step in the potential is not expected to generate 
features in the CMB power spectra of tensor modes. 
Unlike the scalar spectrum whose shape is sensitive to the 
second slow-roll parameter $\eta_V$, the 
shape of the tensor spectrum depends primarily on $\epsilon_V$, which remains small 
at the step (see Fig.~\ref{plot:potential}, \cite{Hamann:2007pa}).  
In Fig.~\ref{plot:tensors}
we show the $B$-mode prediction for $r=0.16$ and a pure power law tensor spectrum with tilt $n_t=-r/8$.

\begin{figure}[tb]
  \resizebox{3.4in}{!}{\includegraphics[angle=0]{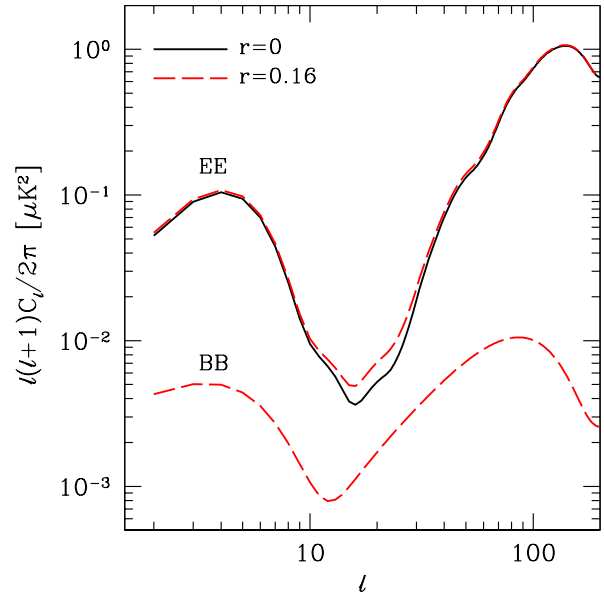}}
\caption{\footnotesize 
Effect of tensor fluctuations on polarization power spectra 
for the model with a feature in $\Pinf(k)$. \emph{Solid black:} no tensor 
component. \emph{Dashed red:} including tensors with $r=0.16$.  Tensors smooth the
$EE$ spectrum near the $\ell \sim 20$ dip.}
\label{plot:tensors}
\end{figure}

On the other hand, it is important to assess the possibility of degradation of the $E$-mode
feature from the curvature spectrum due to the nearly smooth tensor $E$-mode contributions. 
Due to the shape of the tensor $E$-mode spectrum, which mimics the $B$-mode spectrum, the main impact of tensors is to 
fill in the dip in the polarization spectrum around $\ell \sim 20$
(see Fig.~\ref{plot:tensors}).   Correspondingly,
the decrease in significance for the ideal experiment 
is $13-15\%$ in $\sqrt{\dlnl_{EE}}$, and for Planck, $4\%$.
Planck is less affected since its lower sensitivity 
limits the accuracy of measurements in the $\ell\sim 20$ dip.
Since these degradations are relatively small, we ignore tensors
when considering the impact of the reionization history below.   Moreover, we shall
see that reionization uncertainties are very similar to tensors in that they make the
$\ell \sim 20$ dip less useful for distinguishing features through $E$-mode
polarization.

\subsection{Temperature Conditioning}
\label{sec:constrained}

The usefulness of polarization for providing an independent test of features observed in 
temperature may also be reduced by the correlation of temperature and 
polarization: a positive correlation would make observation of 
polarization features more likely given the WMAP $TT$ data 
regardless of whether the features have an inflationary or 
chance statistical origin. We expect the reduction in significance 
to be small given that $C_{\ell}^{TE}$ is small on the relevant 
angular scales (see Fig.~\ref{plot:cltt}), and in this section 
we quantify this statement.

To assess the impact of conditioning polarization predictions on the WMAP temperature 
data, it is convenient to replace the likelihood statistic of Eq.~(\ref{eq:likelihood}) with
a $\chi^2$ statistic.  This allows us to phrase the impact in terms of the bias
and change in variance predicted for the $EE$ power spectrum  from the $TT$ measurements.
Note that in the absence of the temperature constraint and in the limit of small 
differences between the model and the data,
\begin{eqnarray}
\dlnl_{EE} &\approx& f_{\rm sky} \sum_\ell {2 \ell +1 \over 2} 
 {\left(\hat C_\ell^{EE}-  C_\ell^{EE} \over C_\ell^{EE}\right)^2} \nonumber\\
&\approx& \sum_\ell {\left(\hat C_\ell^{EE}-  C_\ell^{EE} \right)^2\over \Var(\hat C_\ell^{EE})},
\label{eqn:chi2}
\end{eqnarray}
which is equal to a simple $\chi^2$ statistic.

Now let us include the temperature constraint.   First take the idealization that the temperature multipole moments  $T_{\ell m}$ have been
measured on the full sky with negligible noise.    Given a model that correlates the 
polarization field through the cross correlation coefficient
\begin{equation}
R_\ell={C_\ell^{TE} \over \sqrt{C_\ell^{TT}C_\ell^{EE}}}\,,
\end{equation}
a constrained realization of the polarization field that is consistent with the temperature
field can be constructed as
\be
{E_{\ell m} \over \sqrt{C_\ell^{EE}}}  = R_\ell {T_{\ell m}  \over \sqrt{C_\ell^{TT}}}
+ \sqrt{1-R_\ell^2}~g_{\ell m}\,,
\ee
where $g_{\ell m}$ is a complex Gaussian field with zero mean, unit variance $\langle g_{\ell m}g_{\ell m}^*\rangle=1$, and a real transform $g_{\ell m}^*=(-1)^m g_{\ell,-m}$.   The estimate of the power spectrum
is then
\begin{equation}
\hat C_\ell^{EE} = {1 \over 2\ell+1} \sum_m E_{\ell m}^* E_{\ell m} \,,
\end{equation}
and its mean over the constrained realizations is biased from the true $C_\ell^{EE}$ \be
{ \langle \hat C_\ell^{EE} \rangle -  C_\ell^{EE} \over C_\ell^{EE} }= R_\ell^2{\hat C_\ell^{TT}- C_\ell^{TT}  \over C_\ell^{TT}} \,,
\ee
by the fixed observed temperature power spectrum $\hat C_\ell^{TT} =\sum_m T_{\ell m}T_{\ell m}^* /(2\ell+1)$.   
With a high correlation coefficient, chance features in the temperature spectrum induce
similar features in the observed polarization spectrum.  For example, 
if $TT$ fluctuates high, $EE$ will also fluctuate high (on average).

The temperature constraint also removes some of the freedom in the variance of the
polarization power spectrum:
\ba
{\Var(\hat C_\ell^{EE}) \over (C_\ell^{EE})^2} &=& {2 \over 2\ell+1}\left(1-R_\ell^2\right)^2 \nn\\
&+&{4\over2\ell+1} R_\ell^2 (1-R_\ell^2){\hat C_\ell^{TT} \over C_\ell^{TT}} \,.
\ea
In the limit that the correlation $R_\ell \rightarrow 0$, the variance takes on its usual form for
a Gaussian random field.   In the limit that $R_\ell \rightarrow 1$, there is no uncorrelated
piece and the observed temperature spectrum determines the observed polarization spectrum
with no variance.

Now let us add in detector noise and finite sky coverage.  Given a noise power spectrum $N_\ell^{EE}$ and a fraction of the sky $f_{\rm sky}$,
\ba
 f_{\rm sky}  {\Var(\hat C_\ell^{EE}) \over(C_\ell^{EE})^2} \approx {2 \over 2\ell+1}\left(1-R_\ell^2 +
{N_\ell^{EE} \over C_\ell^{EE}} \right)^2 \nn\\
+{4\over2\ell+1} R_\ell^2 \left(1-R_\ell^2+ {N_\ell^{EE} \over C_\ell^{EE} }\right){\hat C_\ell^{TT} \over C_\ell^{TT}} \,.
\ea

\begin{figure}[tb]
  \resizebox{3.4in}{!}{\includegraphics[angle=0]{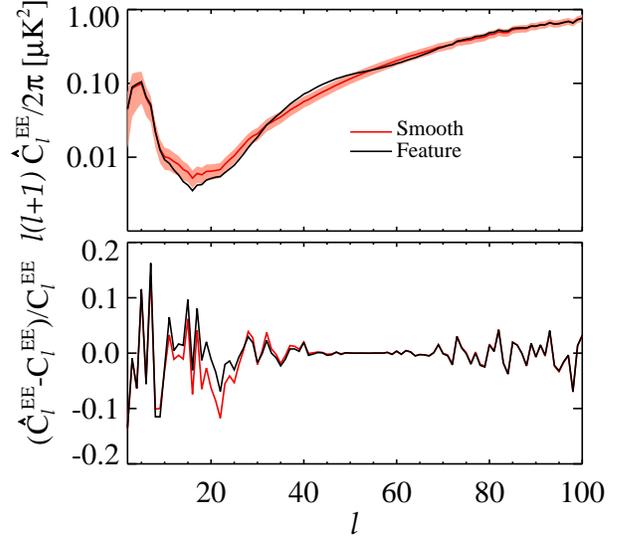}}
\caption{\footnotesize
\emph{Upper panel:}  Solid lines show the
$E$-mode power spectrum constrained to the temperature data for the smooth-$\Pinf(k)$
model along with the band representing sample variance per $\ell$ for the
ideal experiment.  The model with a feature (\emph{dashed}) lies significantly outside of the band in the $10 \simlt \ell \simlt 60$ range, making false negatives unlikely.
\emph{Lower panel:}  Fractional difference between the average
of the constrained realizations $\langle \hat C_\ell^{EE}\rangle$ and the full ensemble average
$C_\ell^{EE}$ for both models.  The impact of the constraint  is minimal due to the lack of correlation between the temperature and polarization fields in the region of interest. 
}
\label{plot:ClEE_constrained_CV}
\end{figure}

Figure~\ref{plot:ClEE_constrained_CV} shows, in the upper panel, the $E$-mode 
polarization power spectrum 
for the smooth inflationary spectrum 
constrained to WMAP5 temperature data for the ideal experiment. 
 For comparison, 
$\clee$ for the best fit feature model  is also plotted.
Note that even the second dip in the spectrum at $\ell \sim 60$ remains 
significantly distinct in polarization.  
In the lower panel, the impact of the temperature power spectrum constraint  
is plotted as the fractional difference between $\hat C_\ell^{EE}$ and $C_\ell^{EE}$ for each model.    Due to the lack of temperature-polarization correlation in the $10 \simlt \ell \simlt 60$
regime, the impact of the constraint on the polarization features is negligible.

We can quantify these conclusions by 
generalizing the $\chi^2$ statistic in Eq.~(\ref{eqn:chi2}) to include
the temperature constraint:
\be
\dchisq_{EE}\equiv \sum_\ell {\left(\hat C_\ell^{EE}- \langle \hat C_\ell^{EE} \rangle \right)^2\over \Var(\hat C_\ell^{EE})} \,.
\ee
As in the likelihood analysis, we assume that the data are a typical draw of the true model (``1") and that we are
testing the significance at which the second model (``2") can be rejected.
Then we set $\hat C_\ell^{EE} = \langle \hat C_\ell^{EE(1)} \rangle$, $\langle \hat C_\ell^{EE} \rangle
= \langle\hat C_\ell^{EE(2)} \rangle$, and Var($\hat C_\ell^{EE}$)= Var($\hat C_\ell^{EE(2)}$).
Note that the bias induced by the temperature constraint enters into both models whereas
the change in the variance enters only from model 2.

\begin{table}
\caption{$\dchisq_{EE}$ for false positive and false negative tests comparing models with smooth $\Pinf(k)$ and a feature in $\Pinf(k)$, with polarization either unconstrained or constrained to observed temperature data.}
\begin{center}
\begin{tabular}{llrr}
\hline
 & & \multicolumn{2}{c}{$\dchisq_{EE}$} \\
Experiment & Test & w/o $T$ & ~with $T$ \\
\hline
\hline
Ideal & False positive &  $70$ & $63$  \\ 
Ideal & False negative &  $59$ & $56$ \\ 
\hline 
Planck & False positive &  $8$ & $8$  \\ 
Planck & False negative &  $9$ & $8$  \\ 
\hline
\end{tabular}
\end{center}
\label{table:instrei2}
\end{table}

Table \ref{table:instrei2} assesses the $\chi^2$ significance of the rejection of false positives and 
false negatives for the fiducial feature model. 
  In the last column we have applied
the constraint from the WMAP5 temperature data and in the penultimate column we
artificially drop the constraint by setting $R_\ell=0$ in the evaluation of $\dchisq_{EE}$.
Even with the constraint, 
the significance with which false positives can be rejected is
$\sqrt{\dchisq_{EE}}=7.9$  for the ideal experiment and $
\sqrt{\dchisq_{EE}}=2.8$ for Planck.
For the case of false negatives, these numbers become
$7.5$ for the ideal experiment and $2.8$ for Planck.   
The difference between false positive and false negative significances
comes from the dependence of sample variance on the model tested.
In all cases, the significance in terms of $\dchisq_{EE}$ is 
comparable to $\dlnl_{EE}$ in Table~\ref{table:instrei1}.

The impact of the temperature constraint is to lower the significance of both
cases but only by $\simlt 5\%$ in $\sqrt{\dchisq_{EE}}$.
This small difference in significance justifies 
our choice to omit the constraint to temperature in our 
exploration of other effects that can degrade the significance.

\section{Reionization Features}
\label{sec:ionization}

A more complicated ionization history can in principle produce features in the polarization
spectrum that might mimic or obscure features from the inflationary power spectrum.
This is especially true for the dip at $\ell \sim 20$.
In this section we search for ionization histories that lead to a higher incidence of
false positives and false negatives.

Reduced significance of false positives or negatives due to confusion 
between features from inflation and from reionization can arise in
two ways. 
First, the true reionization history can
introduce features in the data that either falsely mimic inflationary features
or hide true features.
Second, additional reionization freedom in the (false) model we wish to test
can allow a better match to data generated from the alternate
(true) model assumption.  
To account for both effects, we use a two-step method in which 
we first optimize the ionization history of the true model to produce a 
false positive or negative result, and then vary the ionization history 
of the false model. We will describe this procedure in \S~\ref{sec:optimize}.

Table~\ref{table:reion} summarizes the results of this section.
Relative to the significance of rejecting false positives or
negatives for instantaneous reionization models,
ionization freedom lowers the significance for an ideal experiment by a
factor of $0.64-0.75$ to $\sqrt{\dlnl_{EE}}\approx 5-6$, and
for Planck by a substantially smaller factor of $0.83-0.87$ to $\sqrt{\dlnl_{EE}}\approx 2.5$.
Given the amount of freedom we allow  in the ionization history these should be
viewed as the maximal degradation possible due to reionization.
We describe the details of this calculation in the following sections.

\begin{table}[bt]
\caption{$\dlnl_{EE}$ for tests of false positives and false 
negatives with ionization histories of the data and model tuned 
at $6<z<50$ to minimize the significance of rejection using the methods 
described in \S~\ref{sec:optimize}.
}
\begin{center}
\begin{tabular}{llr}
\hline
Experiment & Test & $\dlnl_{EE}$ \\ 
\hline
\hline
Ideal & False positive & 36 \\ 
Ideal & False negative & 25 \\ 
\hline
Planck & False positive & 6 \\ 
Planck & False negative & 6 \\ 
\hline
\end{tabular}
\end{center}
\label{table:reion}
\end{table}

\subsection{Reionization Principal Components}
\label{sec:pcs}

The form of the ionization history, and therefore the shape of the 
large-scale reionization peak in the polarization spectrum, are only weakly 
constrained by current observations and theoretical modeling, especially 
on the scales relevant for inflationary features \cite{Mortonson:2008rx}. 
We treat the evolution of the mean ionized fraction of hydrogen with redshift, 
$x_e(z)$, as an unconstrained function between $z=6$ and some high 
redshift $z=\zmax$. At lower redshifts, 
we assume $x_e\approx 1$ as required by the observed Ly$\alpha$ transmission 
in quasar spectra at $z\lesssim 6$ (see \eg\ \cite{Fan:2006dp}). The 
highest redshift of reionization is less certain, so we take $\zmax=50$ which 
is quite conservative for conventional sources of ionizing radiation.

We parametrize general reionization histories with a basis of 
principal components (PCs) $S_i(z)$ of the large-scale $E$-mode polarization 
 \cite{Hu:2003gh}.
We use the 7 lowest-variance PCs only since the higher-variance PCs 
have a negligible impact on the polarization power spectrum. Thus 
the ionization history at $6<z<50$ is
\begin{equation}
x_e(z) = x_e^{\rm fid} + \sum_{i=1}^7 m_i S_i(z),
\label{eq:xepc}
\end{equation}
where we take a constant fiducial ionized fraction of $x_e^{\rm fid}=0.07$ 
so that the fiducial model with $\{m_i\}=0$ has a total reionization 
optical depth of $\tau \approx 0.09$.

We vary the PC amplitudes $\{m_i\}$  
and compare the resulting CMB power spectra with data simulated for 
ideal and Planck experiments using Markov Chain Monte Carlo (MCMC) 
likelihood analysis as we describe in the next section.

\subsection{Data and Model Optimization}
\label{sec:optimize}

We use a two-step optimization process to determine the maximum reduction in 
significance of false positive or negative rejection that can be 
caused by reionization features.
We categorize models in this section by whether $\Pinf(k)$ is 
smooth (S) or has a feature (F), and by whether the reionization 
history is instantaneous (I) or more complex (C) and parametrized by 
principal components as in Eq.~(\ref{eq:xepc}).
For comparisons of models we introduce the notation {\tt false}:{\tt true}; 
for example, FI:SI represents the false positive test for instantaneous 
reionization models from \S~\ref{sec:pol_tests}.

In the case of \emph{false positives}, the goal of the optimization 
is to go from 
the FI:SI comparison of the previous section to FC:SC, in which both 
false and true models have complex ionization histories.
In particular, we want to find the FC:SC pair that minimizes the 
difference between the two models, thus minimizing the 
significance of false positive rejection.
To find this optimal pair of models, we use the following procedure:
\begin{enumerate}
\item {\it Optimize true (S) model:} FI:(SI$\to$SC) \\ Vary the ionization history of the SI model to find the SC model that 
best matches FI.
\item {\it Optimize false (F) model:} (FI$\to$FC):SC \\ 
Taking the optimal SC model from step 1 as the true model used to 
generate simulated data, vary the ionization history of 
the FI model to find the FC model that matches the best-fit SC.
\end{enumerate}
Then the significance of rejecting false positives including 
reionization freedom is $\dlnl_{EE}$ computed for the optimal FC:SC
pair, \ie\ the maximum likelihood from step~2.  

These steps for the false positive tests are illustrated in 
Fig.~\ref{plot:reionsteps} for the ideal experiment.
The process for false negatives can be described by simply swapping 
which models have features and which are smooth (F $\leftrightarrow$ S).

Note that even with our conservative choice of $\zmax=50$, 
Fig.~\ref{plot:reionsteps} shows that the main impact of reionization 
on the polarization spectra is limited to $\ell\simlt 30$.
We will explore the consequences of this restriction to large scales 
in the following section.

\begin{figure}[tb]
  \resizebox{3.2in}{!}{\includegraphics[angle=0]{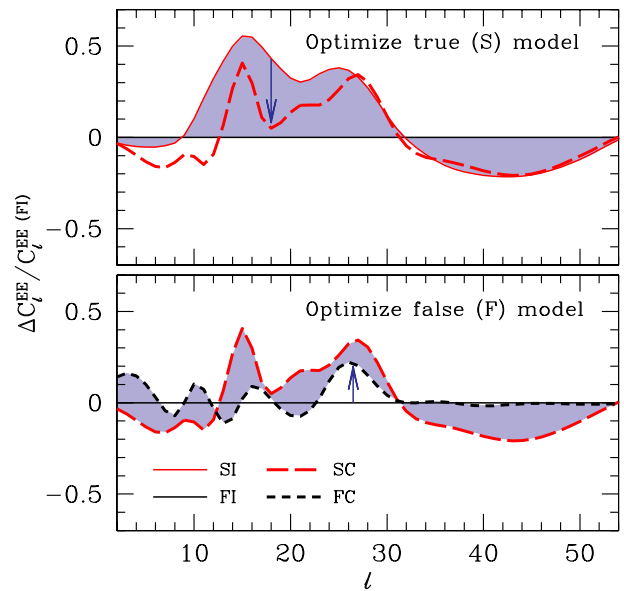}}
  \caption{\footnotesize
False positive example of the two-step process to account for reionization uncertainty in
polarization significance for the ideal experiment.
\emph{Upper panel}: FI:(SI$\to$SC) ---
varying the ionization history 
from SI to SC (\emph{red curves}) 
to match FI (\emph{solid black}). 
\emph{Lower panel}: (FI$\to$FC):SC ---
varying the ionization history from FI to FC
(\emph{black curves}) to match SC (\emph{dashed red}).
All polarization spectra are plotted relative to FI.
This joint optimization minimizes the FI:SI difference 
(\emph{shading in upper panel}, with fiducial significance given in 
Table~\ref{table:instrei1}) at $\ell \simlt 30$
using the optimal FC:SC models (\emph{shading in lower panel}; 
 Table~\ref{table:reion}).
See \S~\ref{sec:optimize} for an explanation of the notation used here.
}
  \label{plot:reionsteps}
\end{figure}

To implement the steps described above, 
we vary ionization histories to minimize 
$\dlnl_{EE}$ following the methods of 
Refs.~\cite{Mortonson:2007hq,Mortonson:2007tb,Mortonson:2008rx},
using CosmoMC \footnote{{\tt http://cosmologist.info/cosmomc/}} 
\cite{Lewis:2002ah} 
for MCMC likelihood analysis with a version of CAMB \cite{Lewis:1999bs} 
modified to include reionization histories parametrized with 
principal components as in Eq.~(\ref{eq:xepc}).
For example, in the FI:(SI$\to$SC) step above, we take the
FI model as the simulated polarization data and search over ionization
histories of the SC model class. 

For each optimization step, we run 4 MCMC chains long enough to be 
well past any initial burn-in phase and stop when the region of 
parameter space near the best fit is sufficiently well sampled 
that all 4 chains agree on the maximum likelihood to within 
$\sim 1\%$ in $\dlnl$. Typically this requires computing an initial chain to 
estimate the covariance matrix of the reionization PC amplitudes, followed 
by generating chains with $\sim 10^4$ samples each.
The optimal true or false model is taken to be the overall
maximum likelihood model from the final 4 chains.

All cosmological parameters besides the 7 
reionization PC amplitudes are assumed to be fixed by measurements of 
the temperature spectrum, except for the amplitude of scalar 
fluctuations $A_s$ which is varied to 
keep $A_s \exp(-2\tau)$ fixed, preserving the temperature and 
polarization power at small scales. Fixed parameters are set to the 
values in Table~\ref{tab:modelparameters}.
We use top-hat priors on the PC amplitudes corresponding 
to $0\leq x_e \leq 1$ as described in Ref.~\cite{Mortonson:2007hq}.
Note that the number of PCs used here is larger than the 3 to 5 needed 
for completeness in Ref.~\cite{Mortonson:2007hq} due to our choice of 
a larger maximum redshift. 

Although we are interested in the ability of polarization data to test 
features appearing in the \emph{observed} temperature spectrum, we include the 
contributions from the \emph{model} $TT$ and $TE$ spectra as well as $EE$ in the likelihood for 
MCMC. Keeping the temperature data in the likelihood ensures that we 
do not obtain models that fit the polarization spectrum well at the 
expense of changing the shape of $C_{\ell}^{TT}$. For example, ionization 
histories with sharp transitions in $x_e$ at high redshift can generate 
polarization power at $\ell\sim 40$ to match inflationary features, but 
these models also add power to the temperature on similar scales 
through an enhanced Doppler effect \cite{Mortonson:2007hq,Mortonson:2008rx}.
For the best-fit models, the contribution of temperature data to the 
likelihood is approximately constant: $\dlnl_{TT}\approx 7$.

\begin{figure}[tb]
  \resizebox{3.2in}{!}{\includegraphics[angle=0]{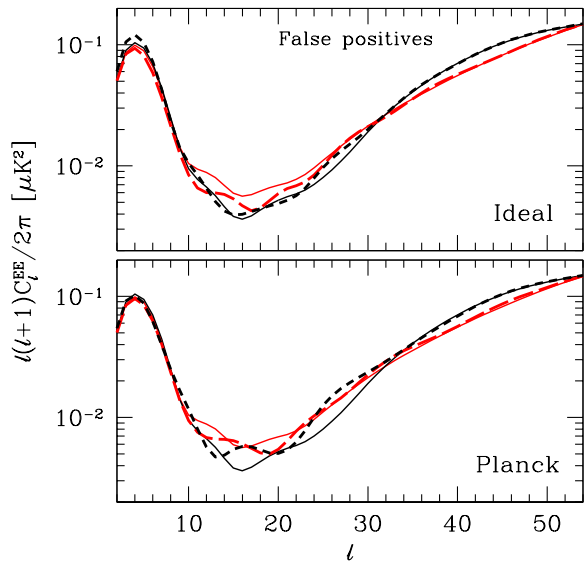}}
  \caption{\footnotesize
Test of false positives due to reionization for an idealized experiment 
limited by sample variance and for Planck.
Thick curves show the true smooth model (\emph{long dashed red}) 
and best-fit false feature model (\emph{short dashed black}) 
for the false positive scenario that would be the most difficult to 
reject due to freedom in the reionization history.
For comparison, the instantaneous reionization polarization spectra 
from Fig.~\ref{plot:cltt} are plotted as thin solid curves.
Reionization histories are parametrized by 7 principal components that 
cover redshifts $6<z<50$.
  } 
  \label{plot:clee1}
\end{figure}

\begin{figure}[tb]
  \resizebox{3.2in}{!}{\includegraphics[angle=0]{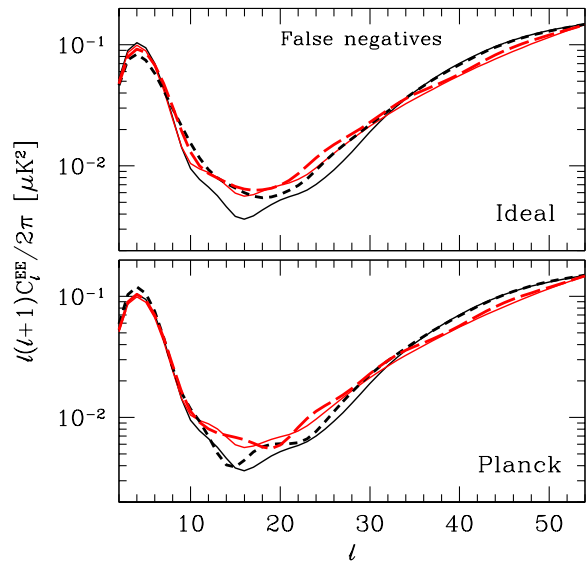}}
  \caption{\footnotesize
Same as Fig.~\ref{plot:clee1}, but for tests of false negatives.
Here the true model has a feature in $\Pinf(k)$ and 
the false model is assumed to have smooth $\Pinf(k)$ ($c=0$).
\vspace{-0.2cm}
  } 
  \label{plot:clee2}
\end{figure}

\subsection{Reionization Confusion}
\label{sec:reionresults}

The optimal true and false model spectra obtained from the steps described in the 
previous section (SC and FC) are plotted in 
Figs.~\ref{plot:clee1} and~\ref{plot:clee2} for each of our 
4 scenarios (ideal/Planck tests of false positives/negatives).
We also plot the corresponding models with 
instantaneous reionization histories (SI and FI)
to show where ionization freedom has the largest effect on 
the spectra.
Table~\ref{table:reion} lists $\dlnl_{EE}$ for each FC:SC or SC:FC
comparison.

The ability of reionization to either mimic or obscure the 
signature of inflationary features is greatest in the low-power 
$10\lesssim\ell\lesssim 30$ regime of $\clee$.
For tests of false positives with an ideal experiment, nearly 
all of the $25\%$ reduction of $\sqrt{\dlnl_{EE}}$ due to reionization 
comes from $\ell<30$. Planck, on the other hand, has relatively greater sensitivity
to small changes in the polarization bump at $\ell\sim 40$ 
since observations at such scales suffer less from instrumental noise 
than at $\ell\sim 20$.
The $17\%$ reduction of $\sqrt{\dlnl_{EE}}$ for false positive rejection for 
Planck due to reionization is split equally between $\ell<30$ and $\ell\ge 30$.
This fact combined with the weakness of reionization effects 
at $\ell\simgt 30$ makes Planck somewhat less sensitive to reionization 
uncertainties than an idealized noise-free experiment.

For false negative tests, changes to polarization spectra at $\ell>30$
are generally more important than they are for false positives.
In the case of the ideal experiment, if we ignored multipoles above $\ell=30$
reionization would only reduce $\sqrt{\dlnl_{EE}}$ by $20\%$ relative to 
the instantaneous reionization significance instead of the $36\%$ 
reduction that we find when including all scales.
The false model being tested in this case (smooth $\Pinf(k)$) has less power and 
therefore lower sample variance at $30\lesssim\ell\lesssim 50$ 
than the spectrum with a feature, and therefore changes in the polarization 
spectra on these scales have a greater effect on the significance 
than they do for false positive tests.
Likewise, Planck's significance is more dependent on the 
$\ell\sim 40$ bump for testing false negatives than for 
false positives. In fact, nearly all of the $13\%$ degradation in 
$\sqrt{\dlnl_{EE}}$ for false negative rejection comes from $\ell>30$ 
for Planck.

Changes in the reionization history at $6<z<50$ are unable to 
significantly affect the polarization power spectrum at 
$\ell \gtrsim 50$. A detection of 
polarization features on these scales would therefore be robust to 
reionization uncertainty. Likewise, measurement of a smooth spectrum 
on these scales would strengthen bounds on the height and width of 
a step in the inflaton potential.

By considering variations in $x_e(z)$ up to $z=50$, 
we include a wide variety of ionization histories, many of which 
may not be physically plausible. In practice, however, the 
ionization histories of the spectra in Figs.~\ref{plot:clee1} 
and~\ref{plot:clee2} have $x_e\simlt 0.2$ at $z>20$.
Nevertheless, had we chosen to limit ionization variation to 
lower redshifts the possibility of confusing reionization with 
inflationary features would be lessened, particularly for 
tests of false negatives and for Planck, due to the greater reliance
on small-scale features in those cases.

Note that the effects of optimizing the ionization history and 
smoothing the polarization spectra with the addition of a large 
tensor component (\S~\ref{sec:tensors}) 
are similar: both are able to make a spectrum with 
inflationary features and a smooth spectrum appear more alike at 
$10\lesssim\ell\lesssim 30$.
Due to this similarity, we expect that considering tensors and 
reionization simultaneously would not further degrade the significance 
of false positive or negative tests.

For variations in the potential parameters discussed in \S~\ref{sec:inflationarydegradation}
 that
retain only the dip at $\ell\sim 20$ and not the bump at $\ell\sim 40$, 
the impact of reionization will be greater.  
In these cases one cannot expect polarization to provide unambiguous 
confirmation of features without external input on the ionization history.

\section{Discussion} \label{sec:discuss}

Models with a step in the inflationary potential produce 
oscillations in the angular power spectra of the CMB that can 
improve  the fit to WMAP temperature data at multipoles $\ell\sim 20-40$ 
at the expense of 3 additional phenomenological parameters controlling the a step 
height, width, and location on the inflaton potential.
Such models predict that these oscillations should appear in the 
$E$-mode polarization spectrum on similar, few-degree scales.
The first precise measurements of the polarization on these scales are 
anticipated in the next few years from Planck, enabling tests of 
the inflationary-step hypothesis.

Moreover, inflationary features at the upper range of $\ell\simgt 30$
($k \simgt 2 \times 10^{-3}$ Mpc$^{-1}$) that are smoothed out 
due to projection effects in temperature should be more visible in 
polarization.
For the lower range of $\ell\simlt 30$ ($k \simlt  2 \times 10^{-3}$ 
Mpc$^{-1}$), it becomes important to assess
the impact of reionization and tensor mode uncertainties.  

We have explored in detail the prospects for polarization tests
of features, focusing 
in particular on the risk of errors that can be classified as
false positives (falsely confirming an inflationary feature) and false negatives (falsely
rejecting an inflationary feature).  Under the simplest set of assumptions for 
large-scale polarization in which we take the best-fit model for the
temperature features, neglect tensor fluctuations, 
and take the reionization history to be instantaneous, 
polarization measurements from Planck should be able to confirm or 
exclude the inflationary features that best match current temperature 
data with a significance of $\sqrt{\dlnl_{EE}} \sim 3$. All-sky 
experiments beyond Planck could potentially increase this significance 
to $\sqrt{\dlnl_{EE}} \sim 8$, providing a definitive test for 
features from inflation. 

The estimated significance degrades slightly with the addition of 
a large-amplitude, smooth tensor component to the $E$-mode spectrum, 
which tends to hide the effect of an inflationary step at the 
largest scales.
Assuming that the step modifies an $m^2\phi^2$ potential, 
$\sqrt{\dlnl_{EE}}$ is reduced by $4\%$ for Planck and $\sim 14\%$ for a 
cosmic variance limited experiment. Allowing non-standard 
reionization histories with arbitrary changes to the ionized fraction 
at $6<z<50$ can lower $\sqrt{\dlnl_{EE}}$ by as much as $\sim 15\%$ for Planck
and $\sim 30\%$ for cosmic variance limited data.
Since tensor fluctuations and reionization have the greatest impact on 
detectability of inflationary features at similar scales ($\ell\sim 20$),
their effects on the significance should not be cumulative.

The possible contamination due to tensors or reionization could eventually 
be mitigated with constraints from other types of observations, 
\eg\ stronger limits on the tensor-to-scalar ratio from the
$B$-mode polarization power spectrum. The $B$-mode contribution
from an $m^2\phi^2$ potential is potentially within the reach of Planck~\cite{Efstathiou:2009kt}.
Note, however, that a failure to reject false positives or 
negatives for inflationary features in $E$-mode polarization would generally
bias the inferred ionization history and reionization parameters 
such as the optical depth.  Such biases would in turn lead to biased 
constraints on inflationary parameters from tensor $B$-mode 
measurements~\cite{Mortonson:2007tb}.

These estimated significances assume that the 
parameters of the step in the inflaton potential are those that best 
fit the WMAP temperature spectrum.  Away from this best fit, the
polarization significance can either increase or decrease.  
Cases where the significance substantially decreases correspond
to parameter combinations where at most one of the dip ($\ell\sim 20$)
and bump ($\ell\sim 40$) temperature features
can be explained by the step in the potential.
In the case that only the dip is inflationary, Planck will be unable to 
confirm the feature.

We have not computed the impact of foreground removal uncertainties on our 
results; in general one might expect our forecasts to degrade somewhat upon 
including them. However, recent studies for Planck \cite{Efstathiou:2009kt} 
and a future dedicated polarization satellite mission 
\cite{Verde:2005ff,Dunkley:2008am} indicate that foregrounds will not be a 
substantial problem in the relevant multipole range.
Finally, we do not address the possibility that the features
in the WMAP data arise from a systematic effect (\cf\ Appendix~\ref{sec:appendix}).
Nonetheless, if all of the $\ell=20-40$ features in the temperature power spectrum
are inflationary, polarization should ultimately provide a statistically significant confirmation.

\vspace{0.6cm}
{\it Acknowledgments:}
We thank Jan Hamann and Richard Easther for valuable conversations.
MJM, CD and WH were 
 supported by the KICP through the grant NSF PHY-0114422 and
the David and
Lucile Packard Foundation. 
MJM was additionally supported through the NSF GRFP. 
WH was additionally supported by  the DOE through 
contract DE-FG02-90ER-40560. HVP was supported in part by 
Marie Curie grant MIRG-CT-2007-203314 from the European 
Commission, and by a STFC Advanced Fellowship. HVP thanks
the Galileo Galilei Institute for Theoretical Physics for the hospitality
and the INFN for partial support during the completion of this work.

\appendix

\section{Relation to Prior Work}
\label{sec:appendix}

Our best-fit parameters for the WMAP temperature data 
(Table~\ref{tab:modelparameters}) differ from those found by 
previous studies of the same inflationary model in 
Refs.~\cite{Covi:2006ci, Hamann:2007pa}. We explain here the reasons 
for these discrepancies, which include the addition of data and 
changes in the likelihood code in going from 3-year to 5-year 
WMAP data, as well as differences in the computation of the 
evolution of modes during inflation.

Due to small changes in the observed $TT$ spectrum between 
WMAP 3-year and 5-year data, we find that the best-fit width of the feature 
increased from $d=0.022$ to a value of $d=0.027$. 
The lower value of $d$ agrees with the best-fit value found by 
Ref.~\cite{Hamann:2007pa}, which was based on the 3-year data.
Note that a wider feature in $k$ implies a narrower feature in $\ell$ 
for the CMB power spectra.
Fig.~\ref{plot:best_fit_models} shows the best fit models for both data sets
along with the appropriately binned data.

\begin{figure}[b]
  \resizebox{3.2in}{!}{\includegraphics[angle=0]{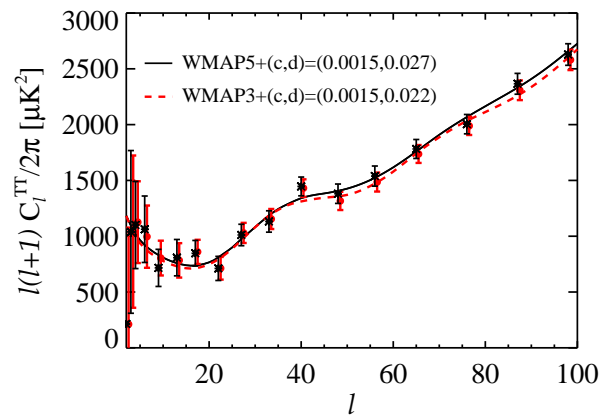}}
\caption{\footnotesize
Observed $TT$ spectrum (binned in $\ell$, with 
WMAP3 and WMAP5 points offset slightly in $\ell$ for clarity) 
and best-fit feature models 
for WMAP3 (\emph{red}) and WMAP5 (\emph{black}). Parameters other than $c$ and $d$ are 
set to the values in Table~\ref{tab:modelparameters} for WMAP5, and 
to the following values for WMAP3: 
$\{m=6.852\times10^{-6}, b=14.67, \Omega_b h^2=0.02222, \Omega_c h^2=0.09927, h=0.753, \tau=0.0817\}$.}
\label{plot:best_fit_models}
\end{figure}

Updates to the WMAP likelihood code could also cause small changes in the 
best-fit potential parameters.  In fact, one might be concerned 
that the feature in the WMAP temperature data is only a systematic 
effect with some artificial origin in the likelihood calculation.
In particular, 
given the location of a feature, its significance could emerge in some
 fashion from the transition between the low-$\ell$ pixel-based $TT$ 
likelihood code and the high-$\ell$ harmonic space likelihood code, which 
happens at $\ell = 32$ in the 5-year likelihood code \cite{Dunkley:2008ie}. 
However, in the original version of the 
3-year likelihood code, {\it v2p1}, 
the transition occurred at $\ell=16$, 
and in the final version, {\it v2p2p2}, it was changed to $\ell = 32$ 
\cite{Hinshaw:2006ia}. 
We searched for the best-fit feature model using WMAP3 
data with these two versions of the likelihood code and found 
almost exactly the same values for the potential parameters
in both cases, indicating that this 
particular issue is not the source of a systematic effect.

Our best fit value for $b$ is considerably different from 
Refs.~\cite{Covi:2006ci, Hamann:2007pa} even though we use the same 
matching condition as they do between $e$-folds and physical wavenumbers. 
This is due to a choice of initial conditions for the background evolution 
of the inflaton by these authors that did not quite satisfy the Friedmann 
equation, with the result that the subsequent evolution also failed to 
satisfy it~\cite{Hamann:pc}. 
This essentially translates into a horizontal shift in $\phi$, 
changing the preferred location of the step $b$.

Ref.~\cite{Hamann:2007pa} discusses relaxing the model dependence of the
predicted power spectrum from the chaotic inflation ``toy model'' adopted 
here by using a free spectral index that is fit to the 
data rather than set by the choice of $N_{\star}$. 
Since the value of $n_s\approx 0.96$ determined by our matching condition 
for the chaotic inflation potential as described in \S~\ref{sec:model}
is nearly identical to the spectral tilt in the WMAP5 best-fit 
concordance model (\ie\ with smooth $\Pinf(k)$), 
we do not carry out this extra step here. 
However, the form of the underlying potential will 
be tested by the Planck satellite irrespective of the existence of features;
as we note in \S~\ref{sec:tensors}, the tensor 
amplitude predicted by the $m^2 \phi^2$ potential (which is not affected 
by the presence of the feature) is within Planck's reach~\cite{Efstathiou:2009kt}.

\bibpreamble{\vspace{-1cm}}
\bibliography{pkfeature}

\end{document}